\documentclass[pdflatex,sn-mathphys-num]{sn-jnl}%

\usepackage{graphicx}%
\usepackage{multirow}%
\usepackage{amsmath,amssymb,amsfonts}%
\usepackage{amsthm}%
\usepackage{mathrsfs}%
\usepackage{bm}
\usepackage[title]{appendix}%
\usepackage{xcolor}%
\usepackage{textcomp}%
\usepackage{manyfoot}%
\usepackage{booktabs}%
\usepackage{algorithm}%
\usepackage{algorithmicx}%
\usepackage{algpseudocode}%
\usepackage{listings}%
\usepackage{setspace}
\begin{document}

\title[In situ local learning of dynamic network materials]{In Situ Local Learning of Dynamic Network Materials}

\author{\fnm{Shuaifeng} \sur{Li}}

\author*{\fnm{Xiaoming} \sur{Mao}}\email{maox@umich.edu}

\affil{\orgdiv{Department of Physics}, \orgname{University of Michigan}, \orgaddress{\city{Ann Arbor}, \postcode{48109}, \state{Michigan}, \country{USA}}}

\abstract{
Functional materials realize functionalities through pre-designed structures, whereas physical neural networks are trained by prescribing input--output behavior. Unifying these views would allow materials to acquire functions directly from dynamical tasks. Here we introduce an in situ local learning framework for dynamic network materials. A forward drive and a time-reversed adjoint error drive are applied to the same mechanical network, enabling the material to compute gradients of a time-domain loss through its own dynamics. The resulting update rules are local, compatible with the principle of physical learning. Each bond or node requires only co-located forward and adjoint fields. The method trains spring constants, nodal masses, bond damping coefficients and nodal damping coefficients, and programs dynamical functions including broadband wave cloaking, evanescent-wave imaging, passive transient output enhancement and vowel classification. These examples involve distinct physical mechanisms, such as wave scattering, near-field information transfer, dissipative non-normal dynamics and machine learning, but are obtained by simply changing the loss function, rather than leveraging expert knowledge. Therefore, dynamic network materials can act both as programmable matter and as physical neural networks, providing a route to material systems that learn temporal responses in situ.
}


\maketitle
\newpage

\section*{Introduction}\label{sec1}
Physical neural networks (PNNs) use the native dynamics of matter to transform physical inputs into outputs, with analogue transformations performed by optical, electronic, chemical, acoustic or mechanical media~\cite{momeni2025training,wetzstein2020inference,sebastian2020memory,lin2018alloptical,wright2022deep}. Recent work distinguishes hardware that is engineered to reproduce a prescribed mathematical operation from broken-isomorphism PNNs, in which the physical transformation itself is trained rather than forced into one-to-one correspondence with a conventional neural-network layer~\cite{momeni2025training,wright2022deep,momeni2023backpropfree}. This viewpoint is especially natural for materials. If local physical parameters can be adjusted from task feedback, a material is no longer only a passive substrate or a fixed device. It becomes an information-processing medium whose function is learned through its own dynamics.

Network materials provide a concrete setting for this idea. Their macroscopic response is encoded in local parameters, stiffness, inertia, connectivity and dissipation, which govern how forces, deformations and waves propagate through the structure~\cite{bertoldi2017flexible,kadic2019metamaterials,ortiz2025unifiedmodellinearresponses}. In the language of PNNs, these quantities are learning degrees of freedom and the response represents the computational results. In the language of mechanics and materials, the same quantities define a design space in which programmable functions and physical optimization objectives can be encoded directly into material parameters. This dual interpretation suggests that a common learning rule could program mechanical functions that range from static shape response and dispersion engineering to transient signal processing and classification, opening a route towards materials with embodied intelligence.

The limiting step is training. The most direct strategy is in silico training, where a differentiable digital twin is trained by gradient descent, and then transferred to the physical hardware~\cite{molesky2018inverse,rumelhart1986learning,wright2022deep,onodera2024scaling}. This approach is powerful, but it shifts the burden to model fidelity. Fabrication disorder, drift, nonlinear contact, damping uncertainty, actuation bandwidth and measurement noise can create a simulation--reality gap between the digital model and the deployed system~\cite{momeni2025training}. Physics-aware training reduces this burden by using the physical hardware for the forward pass and a digital model only for gradient estimation, so that the estimated gradient remains sufficiently aligned with the true gradient~\cite{wright2022deep,pai2023insitu}. Nevertheless, a differentiable representation of the hardware remains part of the training loop.

Local physical learning offers a complementary route to this model-dependent paradigm. Instead of reconstructing a differentiable surrogate of the entire device or communicating gradients through an externally prescribed computational graph, the physical system is asked to supply the information needed for parameter updates through its own fields. In this view, a tunable bond or node should be updated from quantities available where that learning degree of freedom resides, while global task information enters only through boundary conditions, output errors or weak perturbations. Such locality is especially natural for materials, whose trainable parameters are spatially distributed and whose full internal state is costly to measure, store and differentiate digitally. The central challenge then shifts from building an accurate digital twin to designing physical protocols that expose the local correlations required for learning.

Several physical learning schemes already embody this principle. Equilibrium propagation and coupled learning use the contrast between free and weakly nudged states to approximate supervised gradients in energy-minimizing systems, with demonstrations in resistor and elastic networks~\cite{scellier2017equilibrium,stern2021supervised,altman2024experimental,dillavou2024machine}. Frequency propagation extends contrastive ideas to oscillatory physical networks, and Hamiltonian echo backpropagation uses time-reversal symmetry to produce gradient updates through reversible dynamics~\cite{anisetti2022frequency,lopezpastor2023selflearning}. Adjoint-based physical backpropagation provides a forward field and an adjoint error field that are combined locally so that one forward experiment and one adjoint experiment yield gradients with respect to many parameters~\cite{hermans2015trainable,pai2023insitu,plessix2006adjoint,hughes2018training,hughes2019wave,li2024training,li2026topological}. These advances establish that physical systems can participate directly in their own training, but a general in-situ backpropagation rule for arbitrary time-domain functions remains lacking.

Mechanical networks are particularly challenging because their responses depend jointly on elasticity, inertia and dissipation, yet existing backpropagation rules have addressed only restricted regimes. Static in situ backpropagation uses forward and adjoint equilibrium states to obtain exact local gradients for spring constants~\cite{li2024training}. Frequency-domain extensions use harmonic forward and adjoint wave fields to train topological mechanical networks for wave-based classification~\cite{li2026topological}. These methods establish a mechanical analogue of backpropagation, but static learning omits inertia and damping, whereas frequency-domain learning is tied to steady-state harmonic responses at prescribed frequencies. Dynamic network materials instead operate on excitation with various waveforms, with responses shaped jointly by stiffness, mass and damping. A local learning rule for full time-domain mechanics is needed to move from static or steady-state mechanical networks to materials that learn temporal functions.

Here we introduce such a learning framework for dynamic network materials governed by second-order dynamics. A forward input and a time-reversed adjoint error signal are applied to the same physical network, allowing gradients of a time-integrated loss to be computed through the material dynamics. The update rules are local at individual bonds and nodes, and train spring constants, nodal masses, bond damping coefficients and nodal damping coefficients. This generalization allows a single dynamic network-material platform to learn various functions with distinct physical background by only changing the loss function. By making training itself a physical, local and time-domain process, our approach offers a route from inverse-designed network materials to materials that can learn functionalities after fabrication. This capability is crucial for mechanical systems that process information, respond to changing environments and acquire task-specific functions in situ.

\section*{Results}\label{sec2}
\subsection*{In situ local learning rule for dynamic network materials}
We first formulate the local learning rule. Consider a mechanical network of $n$ nodes embedded in $d$ dimensions and connected by $N_b$ linear bonds. The nodal displacement is $u(t)\in \mathbb{R}^{dn}$. The compatibility matrix $C\in \mathbb{R}^{N_b\times dn}$ maps nodal displacements to linear bond elongations, $e(t)=Cu(t)$, with $e(t)\in \mathbb{R}^{N_b}$. The elastic stiffness matrix is $D=C^{T}KC$, where $K=\mathrm{diag}(k_1,\ldots,k_{N_b})$ and $k_j$ is the spring constant of bond $j$. Unlike static mechanical networks, the dynamic network materials considered here also exhibit inertia and dissipation. The mass matrix is $M=\mathrm{diag}(m_1 I_d,\ldots,m_n I_d)$, where $m_i$ is the mass assigned to node $i$ and $I_d$ is the $d$-dimensional identity. Bond damping is $B_{\mathrm{bond}}=C^{T}B_bC$, where $B_b=\mathrm{diag}(b_1,\ldots,b_{N_b})$ and $b_j$ is the damping coefficient of bond $j$. Physically, this term represents viscoelastic loss associated with the strain of a bond. Nodal damping is $B_{\mathrm{node}}=\mathrm{diag}(\beta_1 I_d,\ldots,\beta_n I_d)$, where $\beta_i$ is the drag or friction coefficient of node $i$ with respect to the environment. This term represents local drag or frictional loss acting on the motion of individual nodes. The total damping matrix is $B=B_{\mathrm{bond}}+B_{\mathrm{node}}$. Under an external force $F(t)$, the forward dynamics is
\begin{equation}
    \label{equ1}
    M\ddot u(t)+B\dot u(t)+Du(t)=F(t),
\end{equation}
with prescribed initial conditions. The learning degrees of freedom are the parameter vectors $k$, $m$, $b$ and $\beta$, corresponding to spring constants, nodal masses, bond damping coefficients and nodal damping coefficients, respectively. For a specified target function, we define a time-domain loss
\begin{equation}
    \label{equ2}
    \mathcal{L}=\int_{0}^{T} \ell[u(t),t] dt,
\end{equation}
which measures the mismatch between the actual and desired dynamical responses over $[0,T]$.

Computing the gradient of $\mathcal{L}$ by perturbing each local parameter separately is inefficient, because a perturbation at any bond or node can influence the full network trajectory. We then introduce an adjoint field $u_{\mathrm{adj}}(t)$, following the adjoint-state principle that a full gradient can be obtained from one forward solve and one adjoint solve rather than from separate perturbations of each parameter~\cite{plessix2006adjoint}. The adjoint dynamics satisfy
\begin{equation}
    \label{equ3}
    M\ddot{u}_{\mathrm{adj}}(t)-B\dot{u}_{\mathrm{adj}}(t)+Du_{\mathrm{adj}}(t)=-\left(\frac{\partial \ell}{\partial u}\right)^{T}.
\end{equation}
For the time-integrated loss in Eq.~\eqref{equ2}, the adjoint field satisfies the terminal conditions determined by the boundary terms in the derivation of adjoint equation:
\begin{equation}
    \label{equ4}
    u_{\mathrm{adj}}(T)=0,\qquad
    \dot{u}_{\mathrm{adj}}(T)=0.
\end{equation}
Unlike the forward dynamics, whose state is specified at $t=0$ and evolved toward $T$, the adjoint problem is a final-value problem, whose state is specified at $t=T$, and the adjoint equation is solved from $T$ toward earlier times. Relative to the forward equation, the damping term changes sign as a consequence of the adjoint operation on the first-order time derivative. To rewrite this final-value problem as an ordinary forward-time evolution, we define the time-reversed adjoint field $\hat{u}_{\mathrm{adj}}(t)=u_{\mathrm{adj}}(T-t)$, which gives
\begin{equation}
    \label{equ5}
    M\ddot{\hat u}_{\mathrm{adj}}(t)+B\dot{\hat u}_{\mathrm{adj}}(t)+D\hat u_{\mathrm{adj}}(t)=-\left(\frac{\partial \ell}{\partial u}\Big\vert_{(T-t)}\right)^T.
\end{equation}
After time reversal, the adjoint equation has the same physical operator as the forward equation. This means that the error signal generated during the forward evolution can be recorded, reversed in time, and then re-applied as a driving signal to the same mechanical network. The resulting response implements the adjoint dynamics and the gradients are local:
\begin{equation}
    \label{equ6}
    \begin{split}
    &\frac{\partial \mathcal{L}}{\partial k_j}=\int_0^T e_{\mathrm{adj},j}(t)e_j(t)\,dt, 
    \qquad 
    \frac{\partial \mathcal{L}}{\partial b_j} =\int_0^T e_{\mathrm{adj},j}(t)\dot e_j(t)\,dt,\\
    &\frac{\partial \mathcal{L}}{\partial \beta_i} =\int_0^T \sum_{\alpha=1}^{d}u_{\mathrm{adj},i}^{(\alpha)}(t) \dot u_i^{(\alpha)}(t)\,dt,
    \qquad
    \frac{\partial \mathcal{L}}{\partial m_i}=\int_0^T \sum_{\alpha=1}^{d}u_{\mathrm{adj},i}^{(\alpha)}(t) \ddot u_i^{(\alpha)}(t)\,dt.
    \end{split}
\end{equation}
Bond parameters require only bond-local quantities: forward elongation $e_j$, forward elongation rate $\dot e_j$ and adjoint elongation $e_{\mathrm{adj},j}$. Nodal parameters require only node-local quantities: adjoint displacement $u_{\mathrm{adj},i}$, forward velocity $\dot u_i$ and forward acceleration $\ddot u_i$. The rule is local in the same sense as static mechanical in situ backpropagation, but applies to linear time-domain dynamics. Supplementary Note 1 gives the full derivation.
\begin{figure}[h!]
    \centering
    \includegraphics[width=1\textwidth]{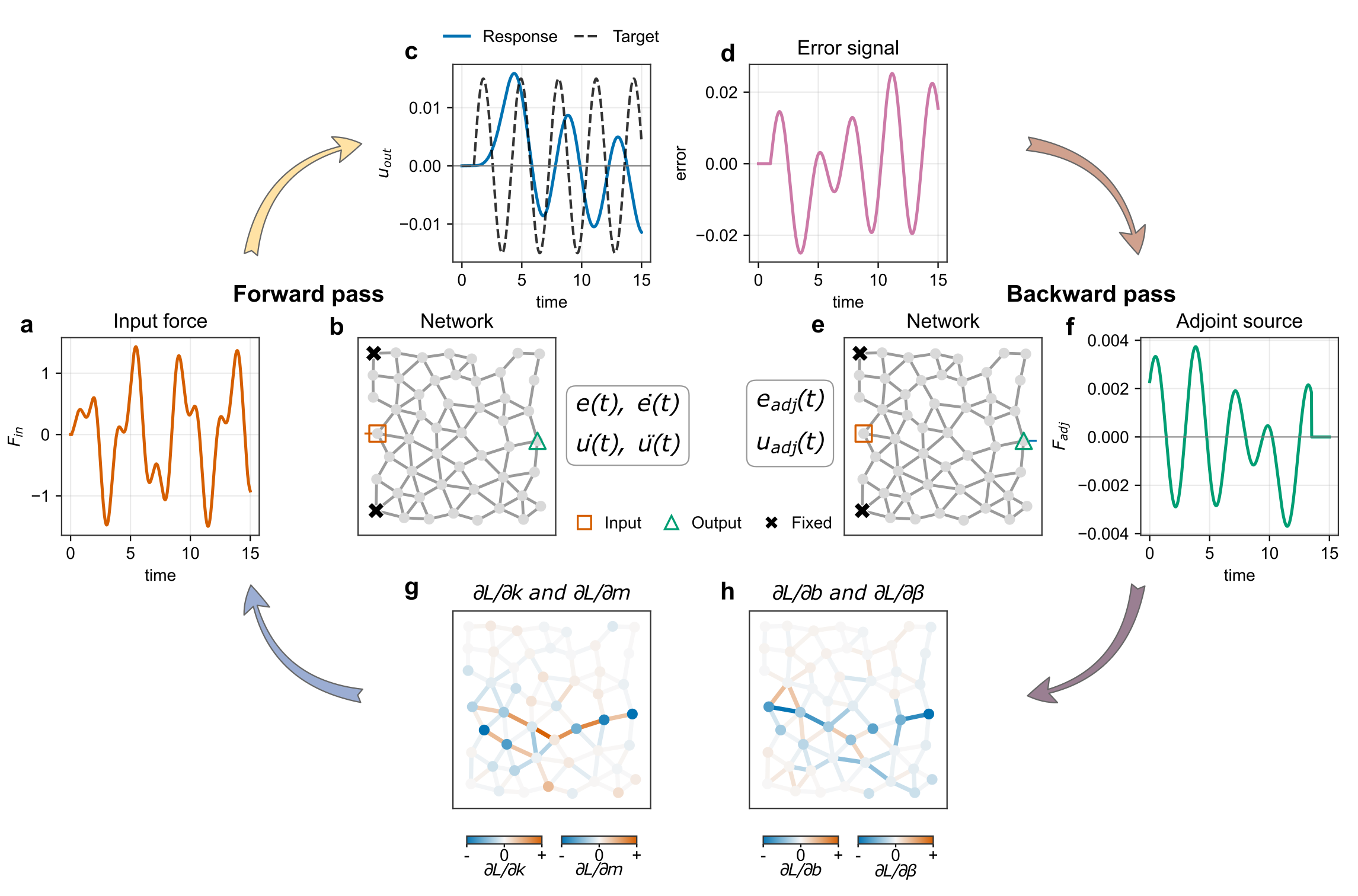}
    \caption{\textbf{Local gradient computation in a dynamic mechanical network.}
    \textbf{a} Input force waveform $F_{\mathrm{in}}(t)$ applied at the input node during the forward pass.
    \textbf{b} Forward-pass network. The orange square marks the driven input node, the green triangle marks the monitored output node and black crosses mark fixed nodes. The forward pass provides $e(t)$, $\dot{e}(t)$, $\dot{u}(t)$ and $\ddot{u}(t)$.
    \textbf{c} Output response and target trajectory. The output displacement $u_{\mathrm{out}}(t)$ is compared with the target waveform $u_{\mathrm{tar}}(t)$.
    \textbf{d} Error signal used to construct the adjoint source.
    \textbf{e} Backward-pass network. The same network is driven at the output degree of freedom by the time-reversed adjoint source, producing $u_{\mathrm{adj}}(t)$ and $e_{\mathrm{adj}}(t)$.
    \textbf{f} Time-reversed adjoint source $F_{\mathrm{adj}}(t)$ applied during the backward pass.
    \textbf{g} Local gradients for spring constants and masses. Bond colors show $\partial\mathcal{L}/\partial k$, and node colors show $\partial\mathcal{L}/\partial m$.
    \textbf{h} Local gradients for bond damping and nodal damping. Bond colors show $\partial\mathcal{L}/\partial b$, and node colors show $\partial\mathcal{L}/\partial\beta$.
    }
    \label{fig1}
\end{figure}

Fig.~\ref{fig1} illustrates the procedure of in situ local learning in a disordered network generated from a jammed-packing contact geometry. The network has one input node on the left, one output node on the right and two fixed boundary nodes that remove rigid-body motion. The input force is applied horizontally, and the output is the projection of the output-node displacement along the same direction.

During the forward pass~(Fig.~\ref{fig1}a, b), the time-dependent force $F_{\mathrm{in}}(t)$ drives the input node and generates the mechanical fields required for learning: $e(t)$, $\dot e(t)$, $\dot u(t)$ and $\ddot u(t)$. The output $u_{\mathrm{out}}(t)$ is compared with a prescribed target $u_{\mathrm{tar}}(t)$~(Fig.~\ref{fig1}c). For response matching, we use
\begin{equation}
    \label{equ7}
    \mathcal{L}=\int_0^T w(t)\left[u_{\mathrm{out}}(t)-u_{\mathrm{tar}}(t)\right]^2 dt,
\end{equation}
where $w(t)$ is nonzero only in the time window of interest. The resulting error signal~(Fig.~\ref{fig1}d) determines the adjoint source.

The backward pass is performed in the same network~(Fig.~\ref{fig1}e, f), but the force is now applied at the output node and equals the time-reversed adjoint source. This experiment produces $u_{\mathrm{adj}}(t)$ and $e_{\mathrm{adj}}(t)$, which are the physical adjoint fields of the network driven by the time-reversed error. Local gradients are then obtained from co-located products of forward and adjoint fields. Fig.~\ref{fig1}g shows gradients with respect to spring constants and nodal masses; Fig.~\ref{fig1}h shows gradients with respect to bond and nodal damping. After obtaining gradient information with respect to each learning degree of freedom, the network can be updated via gradient descent $\theta_{i}\leftarrow \theta_{i}-\eta_{i}\frac{\partial \mathcal{L}}{\partial \theta_{i}}$, where $\theta_{i}$ denotes the $i$-th learning degree of freedom and corresponding learning rate is $\eta_{i}$. Together, these panels show the complete in situ local learning cycle: forward response, output error, time-reversed adjoint response and local gradient readout for all four learning degrees of freedom.

\subsection*{Learning wave cloaking in dynamic network materials}
The demonstration in Fig.~\ref{fig1} establishes that gradients of a time-domain loss can be obtained locally from forward and adjoint fields. We next use this local rule to program wave functions. Rather than prescribing an analytical design, we specify the desired behavior as a target input--output map and update the network parameters from the mismatch between actual and target responses.

We first consider wave cloaking, a canonical problem in wave physics and metamaterials~\cite{pendry2006controlling,leonhardt2006optical,cummer2007path,milton2006cloaking}. A cloak suppresses the scattering signature of an object so that waves outside the cloaked region resemble waves propagating through a reference medium without the object. Conventional approaches include transformation-based designs, scattering cancellation, effective-medium engineering and topology optimization, and typically require an accurate model or a prescribed design ansatz~\cite{pendry2006controlling,cummer2007path,chen2020optimal}. In our formulation, cloaking is formulated as supervised time-domain regression. The trained network must reproduce the response of an obstacle-free background system while the obstacle remains present.

\begin{figure}[h!]
    \centering
    \includegraphics[width=1.0\textwidth]{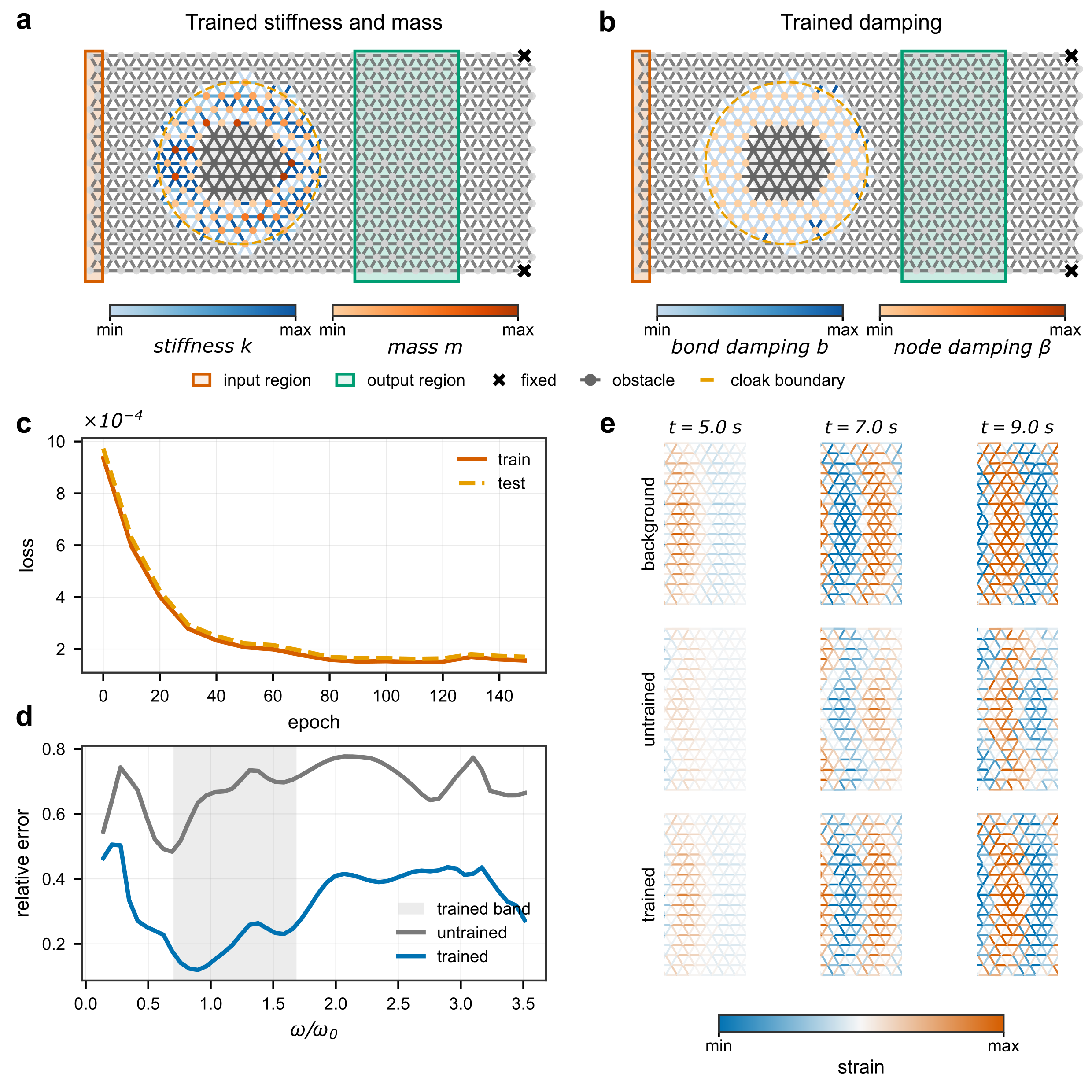}
    \caption{\textbf{Learning broadband wave cloaking over a frequency range.}
    \textbf{a} Trained stiffness and mass distributions in the cloaking network.
    \textbf{b} Trained bond damping and nodal damping distributions in the same network. Panels \textbf{a} and \textbf{b} also show the cloaking geometry: a stiff, heavy obstacle embedded in the network, a trainable annular region, input nodes on the left boundary, downstream output nodes and fixed nodes.
    \textbf{c} Training and testing loss during optimization. Training examples are harmonic incident waves with frequencies sampled from the training band, random phases and reference responses from the obstacle-free background network.
    \textbf{d} Frequency-dependent relative error for the untrained obstacle system and the trained cloaking system. The shaded region marks the training band.
    \textbf{e} Field snapshots in the downstream measurement region for the background network, the untrained obstacle system and the trained cloaking system.
    }
    \label{fig2}
\end{figure}

The setup is shown in Fig.~\ref{fig2}a, b. A two-dimensional triangular network contains a stiff, heavy obstacle that strongly perturbs wave propagation, serving as a representative example of an obstacle that the network learns to accommodate. A surrounding annular region is trainable and serves as the cloak. The remainder of the network is fixed as the background medium. Plane-wave-like excitations are generated by driving a column of input nodes on the left boundary. The response is measured in a downstream strip beyond the obstacle and cloaking region. Selected corner nodes are fixed to remove rigid-body motion, and absorbing sponge layers near the outer boundaries reduce reflections.

To train a broadband cloak, we construct a dataset of incident harmonic waves with frequencies sampled from a finite band. For each example, the frequency is drawn from this band and the phase is randomized. The dataset is randomly divided into training and test sets, containing $70\%$ and $30\%$ of the examples, respectively. Only the training set is used to update the network parameters, while the held-out test set is used to evaluate generalization to unseen frequency–phase combinations. The target is the displacement field produced by the corresponding obstacle-free background network. For the same input, the trainable system contains both the obstacle and the cloaking region. The loss is the mean-squared mismatch between the downstream response of the cloaking system and the reference response,
\begin{equation}
    \label{equ8}
    \mathcal{L}_{\mathrm{cloak}}=\frac{1}{|\Omega_{\mathrm{out}}|}\sum_{i\in\Omega_{\mathrm{out}}}\int_{t_{0}}^{T}|u_{i}(t)-u_{i}^{\mathrm{ref}}(t)|^{2} dt,
\end{equation}
where $\Omega_{\mathrm{out}}$ is the monitored output region, $u_i^{\mathrm{ref}}(t)$ is the obstacle-free response and the initial transient before $t_0$ is excluded. This objective requires that the external dynamical response match the obstacle-free reference over the training band. Training proceeds using mini-batch gradient descent, where at each epoch, a batch of training examples is sampled and our learning rule is used to compute the gradient of the loss.

Figs.~\ref{fig2}a, b show the learned parameters. Stiffness, mass, bond damping and nodal damping vary only in the cloaking region. Training produces heterogeneous distributions around the obstacle that compensate for scattering and reshape the downstream transmitted response. Although damping is trainable, the trained network shows nearly zero damping coefficients.

Training and testing losses decrease during training shown in Fig.~\ref{fig2}c, indicating that the learned cloak generalizes to frequencies and phases not included in a particular training batch. Because training examples are drawn from a frequency band rather than a single harmonic excitation, the network material is trained for broadband response matching within that band. We verify the broadband behavior by sweeping the incident frequency, as shown in Fig.~\ref{fig2}d. Relative to the untrained obstacle system, the trained network reduces the downstream response error across the training band and maintains improved performance over a wider range. Although the training is performed in the time domain, the trained response can be decomposed into frequency components, showing that the time-domain objective produces finite-band wave control rather than a harmonic response at one prescribed frequency.

The field snapshots in Fig.~\ref{fig2}e visualize the effect. In the background system, waves propagate through the measurement region with little distortion. In the untrained obstacle system, the obstacle generates strong scattering and downstream wavefront distortion. After training, the measured field more closely resembles the background response. The trained dynamic network material thus acts as a broadband cloak over the learned band where the obstacle remains present, but its wave signature is suppressed in the monitored region.

This example shows that in situ local learning can program broadband wave control by training learning degree of freedom. Note that the learned cloak is not from a closed-form transformation medium. It is a dynamic network material whose local parameters are trained to match the measured response of a desired reference system.

\subsection*{Learning evanescent-wave imaging}
The cloaking example demonstrates response matching for propagating waves. We next consider a complementary wave function in which the relevant information is carried by near-field spatial structure. Fine object features are encoded in high-spatial-frequency components of the field. Components whose transverse wave number exceeds the propagating threshold are evanescent and decay away from the object, which underlies the diffraction limit of conventional far-field imaging~\cite{pendry2000negative,born1999principles}. Transferring this near-field information is important to subwavelength imaging, near-field sensing and super-resolution wave manipulation.

Evanescent-wave imaging has typically relied on resonant superlenses, negative-index and hyperbolic media with carefully engineered dispersion~\cite{pendry2000negative,liu2007hyperlens,smolyaninov2007magnifying,guo2022evanescent,cheng2018complex}. These approaches are powerful but require a prescribed physical mechanism and geometry. Here, the dynamic network material is trained directly to perform the imaging operation, so it must map near-field object patterns to image-plane responses without imposing a known lens profile.

\begin{figure}[h!]
    \centering
    \includegraphics[width=1.0\textwidth]{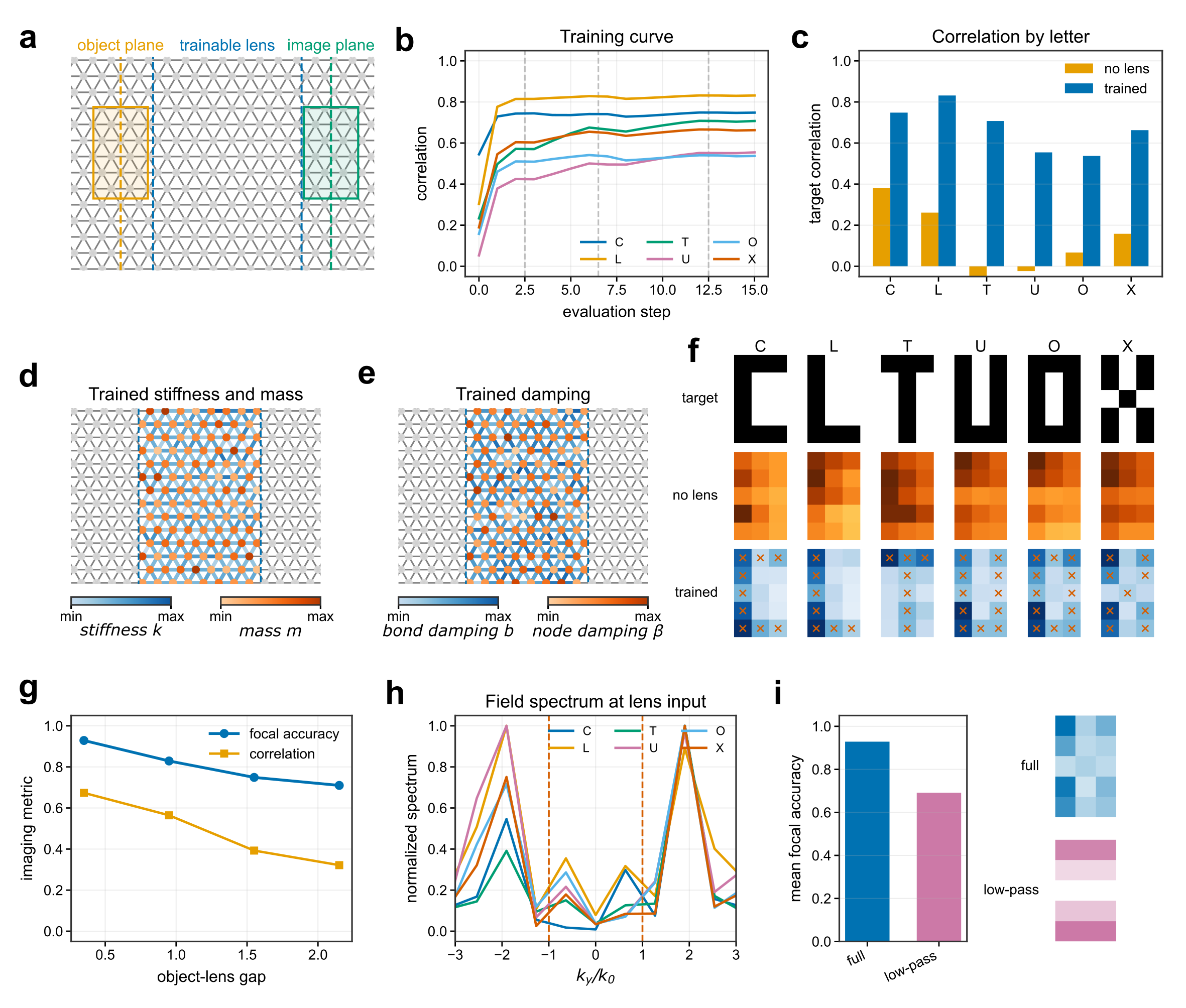}
    \caption{\textbf{Learning evanescent-wave imaging with a dynamic network material.}
    \textbf{a} Imaging task. Six near-field letter targets, $C$, $L$, $T$, $U$, $O$ and $X$, are encoded by active emitting pixels on the object plane. A trainable network region lies between the object and image planes, and the output is measured at corresponding image pixels.
    \textbf{b} Training curve showing target correlation for each letter.
    \textbf{c} Target correlation for each letter with and without the trained lens.
    \textbf{d} Trained stiffness and mass distributions in the lens region. Bond colors denote spring constants $k$; node colors denote nodal masses $m$.
    \textbf{e} Trained damping distributions in the same lens. Bond colors denote bond damping coefficients $b$; node colors denote nodal damping coefficients $\beta$.
    \textbf{f} Rows show target patterns, no-lens responses and trained-lens responses. Orange crosses mark active target pixels in the trained-lens responses.
    \textbf{g} Imaging performance versus object--lens gap. Both focal accuracy and target correlation decrease with gap.
    \textbf{h} Field spectrum at the lens input. Components with $|k_y|/k_0>1$, outside the dashed lines, correspond to evanescent spatial information.
    \textbf{i} Full-spectrum versus low-pass control. The full trained response gives a higher mean focal accuracy and preserves more spatial detail than the propagating-only low-pass response.
    }
    \label{fig3}
\end{figure}

The setup is shown in Fig.~\ref{fig3}a. A triangular mechanical network contains an object plane, a trainable lens region and an image plane. Six letter-shaped targets, $C$, $L$, $T$, $U$, $O$ and $X$, are represented by different patterns of active emitting pixels on the same discrete grid. For each object, active pixels are driven harmonically at a fixed frequency along the in plane $x$ direction, normal to the object and image planes. The resulting near field contains both propagating and evanescent spatial components. The trainable lens is placed near the object plane so that the high-spatial-frequency evanescent components can interact with the network before they decay.

The image is read out from corresponding pixel regions on the image plane. For object $s$, let $P_{i}$ denote the displacement degrees of freedom associated with image pixel $i$. We define the image signal as the time-integrated displacement energy:
\begin{equation}
    \label{equ9}
    r_{i}^{(s)}=\frac{1}{|P_i|}\int_{0}^{T}\sum_{a\in P_i}\left|u_{a}^{(s)}(t)\right|^2 dt.
\end{equation}
$r_{i}^{(s)}$ represents the image produced by the current mechanical lens. We also define a binary target vector $q_i^{(s)}$, with $q_{i}^{(s)}=1$ for pixels that should be active in the desired image and $q_{i}^{(s)}=0$ otherwise. Therefore, $r^{(s)}$ and $q^{(s)}$ provide the measured and desired image patterns used to construct the image-formation loss. For each object, the loss $\mathcal{L}^{(s)}=\mathcal{L}(r^{(s)},q^{(s)})$ encourages the spatial distribution of the measured readout to match the target pattern. Specifically, it promotes comparable responses among the desired pixels while suppressing responses at pixels that should remain inactive.

Importantly, a separate lens is not trained for each letter. A single set of mechanical parameters is trained jointly for all six objects. Conceptually, the image-formation objective function is
\begin{equation}
\label{equ10}
    \mathcal{L}_{\mathrm{img}}
    =
    \sum_{s\in{\{C,L,T,U,O,X\}}}
    w_{s}\mathcal{L}^{(s)},
\end{equation}
where $w_s$ denotes the training weight for object $s$. Thus, at each training step, the response generated by each letter is compared with its corresponding target image, and the resulting losses are combined to update the same lens parameters by gradient descent.

We evaluate image quality using target correlation and focal accuracy. The target correlation is the Pearson correlation coefficient between the normalized image readout $\tilde{r}^{(s)}$ and the binary target pattern $
q^{(s)}$. We also quantify focal accuracy as the fraction of the strongest image-plane responses that occur at the correct target-pixel locations, where the number of selected responses equals the number of active pixels in the target. A focal accuracy of $1$ indicates that all of the strongest reconstructed pixels coincide with the target.

Fig.~\ref{fig3}b displays that target correlations increase during training and then saturate for all six letters. Without the learned lens, image-plane patterns correlate weakly with the targets because near-field information decays before reaching the image plane. After training, the target correlation increases for every letter (Fig.~\ref{fig3}c), demonstrating that the dynamic network material can image the object.

The learned parameter distributions (Fig.~\ref{fig3}d, e) reveal a heterogeneous structure. These elastic, inertial and dissipative parameters define how the near-field response is transformed into an image-plane energy pattern. The reconstructions in Fig.~\ref{fig3}f show that the trained outputs recover the main active-pixel structure of the letters while suppressing other inactive pixels.

We next test whether the learned mechanism depends on evanescent information. Increasing the object--lens gap decreases both the focal accuracy and the target correlation, as shown in Fig.~\ref{fig3}g, as expected because high-spatial-frequency evanescent components decay exponentially away from the object and become less accessible to the lens. The spectral diagnostic in Fig.~\ref{fig3}h provides more direct evidence. Here, $k_{y}$ denotes the transverse spatial wavenumber obtained by Fourier transforming the field along the $y$ direction at the lens input, and $k_{0}=\frac{2\pi}{\lambda_{0}}$ is the reference wavenumber of the homogeneous background at the driving frequency. The measured spectrum contains substantial weight for $|k_y|/k_0>1$, showing that high-spatial-frequency near-field information reaches the lens. These component encodes spatial variation finer than those carried by the propagating part of the field. Finally, we remove the high-spatial-frequency content from the trained response and retain only the propagating low-pass contribution. This control substantially degrades the reconstructed images and reduces focal accuracy~(Fig.~\ref{fig3}i). Therefore, the gap dependence, spectral content and low-pass comparison show that the trained network exploits near-field evanescent components for imaging.

This example demonstrates that in situ local learning can program a dynamic network material to manipulate information inaccessible to ordinary far-field propagation. Whereas the cloaking task suppresses the external signature of an obstacle, evanescent-wave imaging requires the material to preserve and transform high-spatial-frequency object information. The same learning rule hence programs qualitatively different wave functions by changing the target input--output map.

\subsection*{Learning passive transient output enhancement}
The cloaking and imaging tasks show that dynamic network materials can learn wave transformations through their input--output response. We next consider a function controlled more directly by dissipation. In passive mechanical systems, damping usually attenuates motion. When damping is spatially nonuniform and nonproportional to stiffness and mass, however, it can reorganize modal interactions and strongly reshape transient responses before eventual decay~\cite{trefethen1993hydrodynamic,trefethen2005spectra,schmid2007nonmodal}. Such passive transient enhancement is relevant to signal detection, vibration sensing, energy localization and mechanical triggering. Hence, we ask whether our local learning can redistribute a fixed passive damping budget to enhance a desired transient response without introducing active gain.

\begin{figure}[h!]
    \centering
    \includegraphics[width=1.0\textwidth]{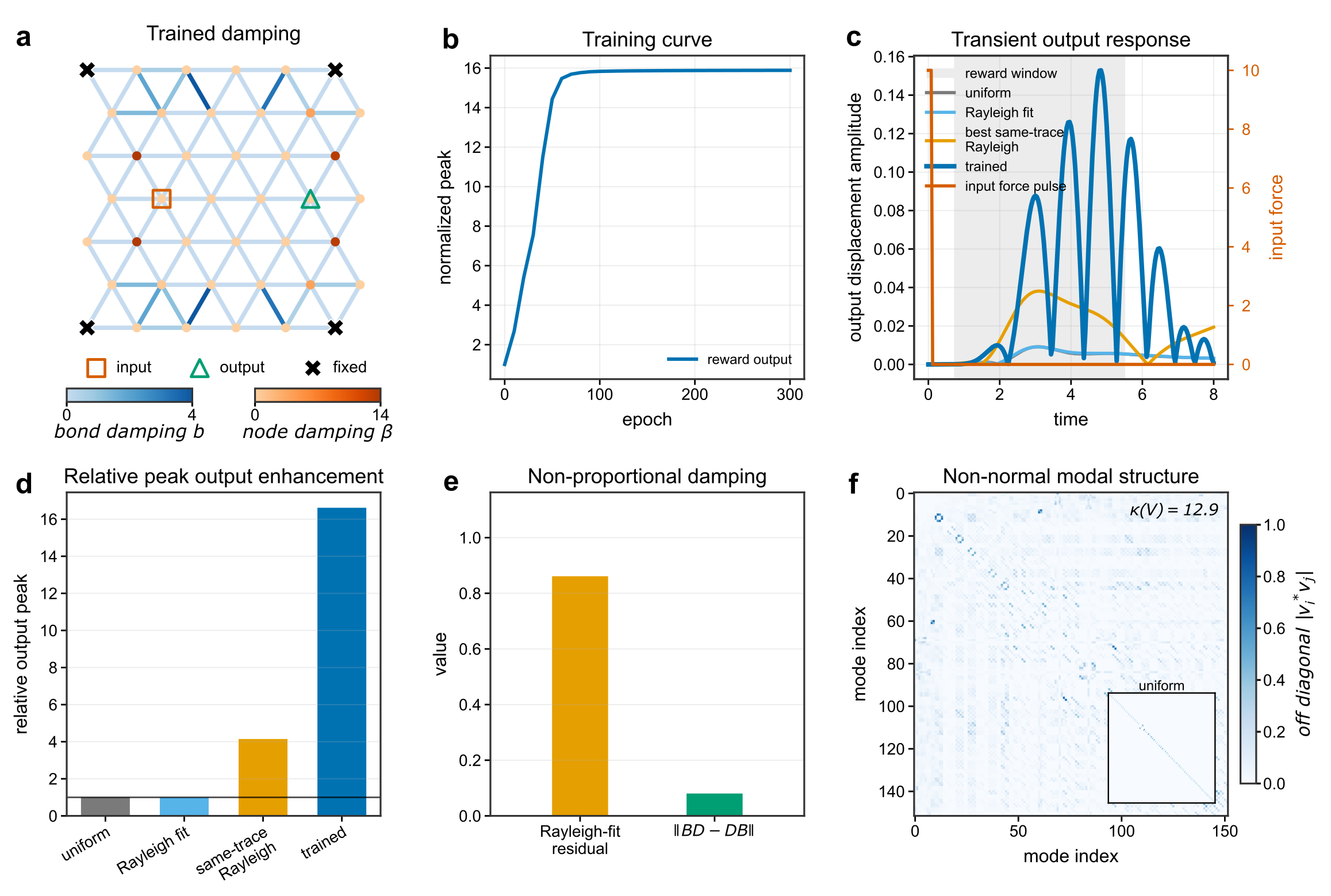}
    \caption{\textbf{Learning passive transient output enhancement by damping redistribution.}
    \textbf{a} Trained damping distribution. A short force pulse with fixed integrated impulse is applied at the input node, and displacement is measured at a distant output node. Spring constants and nodal masses are fixed. Only bond damping $b$ and nodal damping $\beta$ are trained. The total bond- and nodal-damping budgets are fixed and all damping coefficients remain positive.
    \textbf{b} Reward-window peak output displacement during training, normalized by its initial value.
    \textbf{c} Transient output response of the trained network and passive controls. The shaded region marks the reward window. Note that curves for the uniform and Rayleigh-fit networks are overlapped.
    \textbf{d} Relative peak output enhancement, defined as the reward-window peak output displacement normalized by that of the uniform-damping reference under the same excitation.
    \textbf{e} Nonproportional damping diagnostics. The Rayleigh-fit residual and normalized commutator $\|BD-DB\|$ quantify departure from proportional damping.
    \textbf{f} Non-normal modal structure of the trained state-space dynamics. Off-diagonal eigenvector overlaps reveal modal nonorthogonality. The condition number is $\kappa(V)=12.9$. The uniform-damping case is shown as an inset. 
    }
    \label{fig4}
\end{figure}

We train the mechanical network shown in Fig.~\ref{fig4}a to produce a large transient displacement at a prescribed output node following excitation at an input node. In this task, the spring constants and nodal masses are fixed, and only the bond damping coefficients $b$ and nodal damping coefficients $\beta$ are trained. To enforce passivity and enable controlled comparison, every damping coefficient remains positive and the total bond-damping and node-damping budgets are held fixed throughout training. Learning redistributes an existing passive damping budget rather than adding damping or active gain.

A short force pulse with fixed integrated impulse is applied at the input node. Let $a_{\mathrm{out}}(t)=\|P_{\mathrm{out}}u(t)\|_2$ and $a_{\mathrm{in}}(t)=\|P_{\mathrm{in}}u(t)\|_2$ denote the displacement amplitudes at the output and input nodes, respectively, where $P_{\mathrm{out}}$ and $P_{\mathrm{in}}$ project the full displacement field onto the corresponding degrees of freedom. We train the damping distribution to produce a large output transient within a prescribed reward window $\mathcal{T}_{\mathrm{R}}$, while suppressing late output motion, input-side reflections and residual late-time mechanical energy. The resulting objective function is
\begin{equation}
    \label{equ11}
    \begin{split}
    \mathcal{L}_{\mathrm{TA}}=&-\mathcal{S}_{\tau}\left[a_{\mathrm{out}}^2(t);\mathcal{T}_{\mathrm{R}}\right]\\
    &+\lambda_{\mathrm{out}}\int_{\mathcal{T}_{\mathrm{late}}}a_{\mathrm{out}}^2(t)dt
    +\lambda_{\mathrm{in}}\int_{\mathcal{T}_{\mathrm{refl}}}a_{\mathrm{in}}^2(t)dt
    +\lambda_{\mathrm{state}}\int_{\mathcal{T}_{\mathrm{late}}}E(t)dt,
    \end{split}
\end{equation}
where $E(t)=\frac{1}{2}\dot u^T M\dot u+\frac{1}{2}u^T D u$ is the mechanical state energy. The first term rewards a large output response during the desired time interval, with $\mathcal{S}_{\tau}$ denoting a differentiable soft maximum of $a_{\mathrm{out}}^{2}(t)$ over the reward window $\mathcal{T}_{\mathrm{R}}$, where $\tau$ controls how closely it approximates the true maximum, while the remaining terms discourage persistent ringing and reflected or residual motion.

The trained damping distribution is shown in Fig.~\ref{fig4}a. Because the total damping budgets are fixed, the heterogeneous pattern produced by training directly reflects a spatial redistribution of dissipation. During training, the peak output displacement within the reward window increases substantially relative to its initial value~(Fig.~\ref{fig4}b).

To determine whether the enhancement can be reproduced by conventional proportional damping, we compare the trained network with two Rayleigh damping controls~\cite{caughey1960classical,clough1993dynamics}. In mass-normalized free coordinates, a proportional Rayleigh model has $B_{\mathrm{R}}(\eta_D,\eta_M)=\eta_DD+\eta_MI$, where $D$ is the mass-normalized stiffness operator, $I$ is the identity and $\eta_D,\eta_M\ge0$. The Rayleigh-fit control is a nonnegative least-squares approximation to the trained damping matrix $B_{\mathrm{train}}$, rescaled to match $\mathrm{Tr}[B_{\mathrm{train}}]$. A stronger control is obtained by directly selecting the same-trace Rayleigh model that maximizes the reward-window output peak,
\begin{equation}
    \label{equ12}
    \begin{split}
    (\eta_D^\star,\eta_M^\star)=&
    \underset{\eta_D,\eta_M\ge0, t\in\mathcal{T}_{\mathrm{R}}}{\mathrm{argmax}}\left\|P_{\mathrm{out}}u_{\mathrm{R}}(t;\eta_D,\eta_M)\right\|_2,\\
    &\mathrm{subject\ to}\quad
    \eta_D\,\mathrm{Tr}[D]+\eta_M n_f=\mathrm{Tr}[B_{\mathrm{train}}],
    \end{split}
\end{equation}
This provides a proportional-damping comparator with the same total damping trace as the trained system.

Fig.~\ref{fig4}c compares the complete transient output response for uniform damping, the Rayleigh fit, the best same-trace Rayleigh control and the trained damping distribution. Although all systems are passive and are subjected to the same input force pulse, the uniform-damping network produces a much weaker response at the target output. The Rayleigh-fit model also fails to reproduce the enhanced transient response whose curve overlaps with that of the uniform-damping network, showing that simply matching the overall damping strength of the trained system is insufficient. Even the Rayleigh model optimized directly for the largest output peak under the same damping trace remains below the trained network.

We quantify the enhancement relative to the uniform passive reference by
\begin{equation}
    \label{equ13}
    H_q=
    \frac{
    \displaystyle\max_{t\in\mathcal{T}_{\mathrm{R}}}
    a_{\mathrm{out}}^{(q)}(t)}
    {
    \displaystyle\max_{t\in\mathcal{T}_{\mathrm{R}}}
    a_{\mathrm{out}}^{(\mathrm{uniform})}(t)} ,
\end{equation}
where $q$ denotes the damping model. Thus, $H_q>1$ indicates a larger transient output than that of the uniform-damping network under the same excitation. The trained network gives the largest relative peak output enhancement among the tested passive models~(Fig.~\ref{fig4}d), showing that the spatial organization of damping substantially affects the transient response.

To characterize this organization, we quantify the departure from proportional damping. The Rayleigh-fit residual is $\epsilon_{\mathrm{R}}=\min_{\eta_D,\eta_M\ge0}\frac{\|B-B_{\mathrm{R}}\|_F}{\|B\|_F}$, and the normalized commutator is $\chi_{\mathrm{BD}}=\frac{\|BD-DB\|_F}{\|B\|_F\|D\|_F}$. Both quantities are finite for the trained network, as shown in Fig.~\ref{fig4}e, demonstrating that learning produces a nonproportional damping matrix that cannot be represented accurately by a conventional Rayleigh form.

The resulting modal structure provides a dynamical interpretation of the transient enhancement. Writing the second-order system in first-order form gives:
\begin{equation}
    \label{equ14}
    A=\begin{bmatrix}
        0 & I \\
        -M^{-1}D & -M^{-1}B
    \end{bmatrix}.
\end{equation}
If $AV=V\Lambda$, with normalized eigenvectors $v_p$ as columns of $V$, the off-diagonal modal-overlap matrix is $O_{pq}=|v_p^\dagger v_q|$ for $p\ne q$. Large overlaps indicate nonorthogonal eigenvectors and non-normal dynamics. The trained damping distribution produces stronger modal nonorthogonality than uniform damping~(Fig.~\ref{fig4}f), allowing decaying modes to interfere constructively over a finite time interval. This non-normal modal structure provides a passive mechanism for the enhanced transient output without introducing active gain.

This example shows that damping can serve not only as a loss mechanism but also as a trainable resource for shaping transient dynamics. By redistributing a fixed passive damping budget, in situ local learning discovers a nonproportional damping architecture that strongly enhances the transient output relative to uniform and proportional-damping references. The same framework also programs other canonical wave-control objectives. Supplementary Figure 1 shows broadband mechanical focusing, Supplementary Figure 2 shows coherent perfect absorption and Supplementary Figure 3 shows retraining from a focusing material to an absorbing material. We next turn to a data-driven task in which the desired function is specified by class labels rather than a target physical field.

\subsection*{Vowel classification with dynamic network materials}
The preceding examples show that in situ local learning can program physical functions including wave cloaking, evanescent-wave imaging and passive transient output enhancement. We finally apply the same principle to temporal classification, returning to the viewpoint of physical neural networks. Here the mechanical network maps time-dependent inputs to discrete output classes.

We use vowel classification as a representative acoustic task. Vowels are distinguished by temporal waveforms and spectral features associated with vocal-tract resonances~\cite{hillenbrand1995acoustic,hughes2019wave}. Conventional machine-learning pipelines process such information digitally through feature extraction and classification. Here the dynamic network material performs the transformation physically. Audio waveforms from three vowel classes, /iy/, /ae/ and /ei/, are encoded as time-dependent mechanical inputs. The response is read out from three output regions, one assigned to each class.

For input waveform $s$, the readout for class $c$ is the accumulated output energy in region $\Omega_c$,
\begin{equation}
    \label{equ15}
    R_c^{(s)}=\int_0^T\sum_{i\in\Omega_c}\sum_{\alpha}\left|u_{i\alpha}^{(s)}(t)\right|^2 dt.
\end{equation}
Class logits are logarithmic energies, $z_c^{(s)} =\log\left(R_c^{(s)}+\epsilon\right)$, and class probabilities are $p_c^{(s)}=\frac{\exp z_c^{(s)}}{\sum_{c'}\exp z_{c'}^{(s)}}$. For one-hot labels $y_c^{(s)}$, the classification loss is
\begin{equation}
    \label{equ16}
    \mathcal{L}_{\mathrm{vowel}}=-\sum_s\sum_cy_c^{(s)}\log p_c^{(s)}.
\end{equation}
Training encourages each waveform to concentrate time-integrated mechanical energy in the output region corresponding to its label. The predicted class is the output region with the largest accumulated energy. We randomly split the dataset into training set~($70\%$) and testing set~($30\%$). Classification accuracy is defined as the fraction of samples for which the output channel with the largest response corresponds to the correct vowel class. We will report this quantity separately for the training and held-out test sets later.

\begin{figure}[h!]
    \centering
    \includegraphics[width=1.0\textwidth]{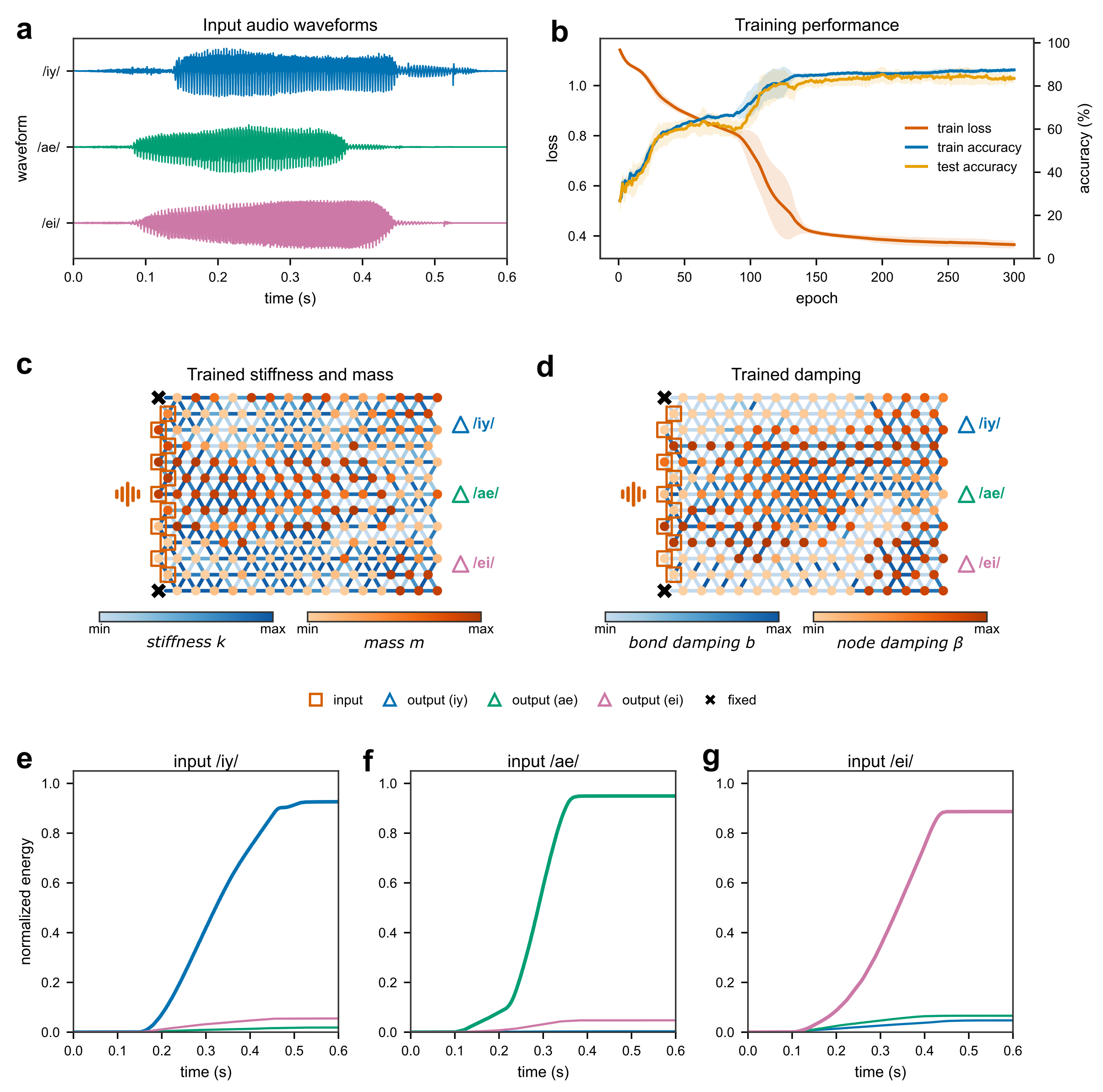}
    \caption{\textbf{Vowel classification with a learned dynamic network material.}
    \textbf{a} Representative input audio waveforms for the /iy/, /ae/ and /ei/ classes. Acoustic signals are encoded as time-dependent mechanical inputs.
    \textbf{b} Training performance. Loss decreases while training and test accuracies increase and saturate. Shaded regions indicate variation across independent training runs or data splits.
    \textbf{c} Trained stiffness and mass distributions. Bond colors denote spring constants $k$; node colors denote nodal masses $m$. Orange squares indicate input nodes, colored triangles indicate class output regions and black crosses indicate fixed nodes.
    \textbf{d} Trained damping distributions. Bond colors denote bond damping coefficients $b$; node colors denote nodal damping coefficients $\beta$.
    \textbf{e--g} Cumulative normalized output energies for representative inputs from the three vowel classes. The correct output channel accumulates the largest energy for /iy/ in \textbf{e}, /ae/ in \textbf{f} and /ei/ in \textbf{g}.
    }
    \label{fig5}
\end{figure}

Representative input waveforms are shown in Fig.~\ref{fig5}a. The signals differ in envelope and oscillatory content, and the relevant information is distributed over time rather than encoded in a single instantaneous displacement. During training, the material parameters are updated so that each vowel selectively excites the output region associated with its class.

Loss decreases during training, while training and testing accuracies converge to approximately $88\%$ and $86\%$, respectively (Fig.~\ref{fig5}b). The trained network implements a physical transformation that generalizes beyond the training examples. The shaded regions show variation across independent runs and data splits.

Figs.~\ref{fig5}c, d show the learned material parameters. The input nodes lie along the left boundary, and three output regions on the right boundary correspond to the vowel classes. Training produces heterogeneous stiffness, mass and damping distributions that define the mechanical transformation from acoustic waveforms to class-selective output responses.

The cumulative normalized output energies for representative inputs are shown in Figs.~\ref{fig5}e--g. For an /iy/ input, the /iy/ channel accumulates the largest energy; for /ae/, the /ae/ channel dominates; and for /ei/, the /ei/ channel becomes largest. Separation between correct and incorrect channels increases over time as the network integrates temporal information from the waveform.

This task shows that locally trained dynamic network materials can operate as physical neural networks for temporal signals. The same class of trainable mechanical systems that realizes wave cloaking, evanescent-wave imaging and passive transient enhancement also performs supervised classification when the loss and readout are changed. The demonstrations support a unified view of dynamic network materials as locally trainable physical systems for both material functionality and machine-learning tasks.

\section*{Discussion and Conclusion}
In this work, we introduce an in situ local learning framework for dynamic mechanical network materials. The central result is that a network material can compute gradients of a time-domain loss through its own physical dynamics. A forward experiment measures the response to an input signal and an adjoint experiment with a time-reversed error signal. Local products of forward and adjoint fields give gradients for spring constants, nodal masses, bond damping coefficients and nodal damping coefficients. This extends mechanical in situ backpropagation from static equilibrium and steady-state harmonic response to general time-domain dynamics.

The framework links physical neural networks and programmable network materials. From the neural-network perspective, the mechanical system performs analog temporal computation, with trainable material parameters serving as learning degrees of freedom. From the materials perspective, the same parameters encode a target function in the network response. The learned system is therefore both a computational device and a dynamic material whose wave propagation, transient response and input--output behavior are programmable by local learning.

We demonstrate this principle across representative tasks. The response-matching example validates the local learning rule for different trainable degrees of freedom. Wave cloaking and evanescent-wave imaging show that the method can learn wave-control functions by matching desired physical input--output maps. Passive transient output enhancement demonstrates that damping can be trained to organize non-normal dynamics and transiently concentrate response at a target output while preserving passivity. Vowel classification shows that the same material can process temporal data and perform supervised classification. Multifunctionality in these examples does not require separate task-specific design rules and expert knowledge. It arises from a common physical learning rule applied to different losses. Beyond training separate networks for different tasks, the same learned material can be reprogrammed by continued local learning. Supplementary Note 4 and Supplementary Figure 3 show that a focusing network can be retrained into a coherent-perfect-absorption network by changing only the objective.

Locality is a key feature of the method. Each gradient depends only on co-located dynamical fields, which is essential for future physical implementations in which centralized measurement and computation of all sensitivities would be impractical~\cite{momeni2025training}. In principle, local sensors, memory elements and adaptive mechanical components could update material parameters from measured forward and adjoint fields.

The framework also clarifies the role of damping in dynamic learning. Damping is not merely a parasitic loss, while, depending on the objective, it can be incidental, beneficial or essential. In cloaking, the learned response is dominated by conservative stiffness and inertial structure. In evanescent-wave imaging and passive transient output enhancement, the learned damping distribution is significant to modulating near-field response and shaping nonproportional modal coupling. This task-dependent use of damping distinguishes dynamic network materials from purely conservative designs and suggests broader opportunities for programming dissipation as a design degree of freedom, even under passive constraints.

Several challenges remain before autonomous experimental realization. The present demonstrations assume access to dynamical fields required for local updates, including displacements, velocities, accelerations, bond elongations and elongation rates. In hardware, these quantities must be measured, inferred or encoded locally, and future implementations will need to tolerate hardware noise, copy-to-copy variation and drift~\cite{momeni2025training}. The trainable parameters must also be physically adjustable. Mechanically reconfigurable metamaterials, magnetorheological and hard-magnetic composites, pneumatic or vacuum-jamming elements, electroactive elastomers and fluid-solid architectures provide established routes for tuning effective stiffness, inertia and dissipation~\cite{florijn2014programmable,rafsanjani2019programming,montgomery2021magnetomechanical,brown2010universal,rizzello2017robust,zhang2018tunable,jolly1996model}, but each implementation will constrain range, resolution, speed and reversibility. Finally, although adjoint dynamics can in principle be generated by time-reversed error signals in the same network, experiments must contend with noise, finite bandwidth, actuation limits and imperfect boundary conditions. These factors will determine the accuracy and stability of physical training.

Scaling is another important direction. Adjoint-based learning is favorable for large systems because a full gradient requires one forward pass and one adjoint pass, independent of the number of trainable parameters~\cite{plessix2006adjoint}. Larger networks, however, introduce longer time windows, richer modal spectra, stronger parameter degeneracy and more complex loss landscapes. Thus, understanding sloppiness, robustness and generalization in learned dynamic materials will be important.

The approach is not limited to the linear mechanical networks studied here. The same adjoint-learning principle applies to physical systems governed by differentiable dynamical equations, including nonlinear mechanical networks, acoustic and elastic continua, electrical and fluidic networks, and hybrid electromechanical systems. Nonlinearity would enable richer computations and amplitude-dependent material functions, while requiring adjoint formulations with state-dependent operators.

In summary, dynamic network materials can be trained locally and in situ by physically backpropagating time-reversed error signals. A single learning rule programs wave cloaking, evanescent-wave imaging, passive transient output enhancement and vowel classification. By allowing stiffness, inertia and damping to act as learning degrees of freedom, the framework provides a route to network materials that learn temporal responses through their own dynamics.

The authors acknowledge National Science Foundation Center for Complex Particle Systems~(Award No. 2243104). The authors would like to thank Prof. Jennifer Schwarz for the insightful discussion.

\backmatter

\bibliography{references}

\end{document}